\documentclass[%
 aip,
 jcp,%
 amsmath,amssymb,
 reprint,%
a4paper
]{revtex4-1}

\usepackage[english]{babel}
\usepackage[utf8x]{inputenc}
\usepackage{array}
\usepackage{graphicx}
\usepackage[version=3]{mhchem}

\newcolumntype{P}[1]{>{\centering\arraybackslash}p{#1}}

\begin{document}
	\title{Thermal laser evaporation of elements from across the periodic table}
	
	\author{Thomas J. Smart}
	\author{Jochen Mannhart}
	\author{Wolfgang Braun}
	\email[Corresponding Author: ]{w.braun@fkf.mpg.de}
	
	\affiliation{Max Planck Institute for Solid State Research, \\
		Heisenbergstraße 1, 70569, Stuttgart}%
	\date{\today}
\begin{abstract}
We propose and demonstrate that thermal laser evaporation can be applied to all solid, non-radioactive elements in the periodic table.
By depositing thin films, we achieve growth rates exceeding 1\,Å/s  with output laser powers less than 500\,W, using identical beam parameters for many different elements.
The source temperature is found to vary linearly with laser power within the examined power range.
High growth rates are possible using free-standing sources for most of the elements tested, eliminating the need for crucibles. 
\end{abstract}
\keywords{thermal laser evaporation, film growth, epitaxy, thin films, deposition}
\maketitle
Shortly after the invention of the laser in 1960\cite{Maiman_1960}, attempts were made to synthesize thin films by using a laser to evaporate or ablate materials.\cite{Smith:65, Groh}
These initial experiments would go on to lay the groundwork for Pulsed Laser Deposition (PLD):\cite{PLDWillmott} a successful vapor deposition technique that ablates the surface of a source using a pulsed laser.\cite{Schwarz,pld}
The advantages of using lasers for epitaxial growth are numerous, including near arbitrary power densities, lack of source contamination and increased efficiency due to the surface of the source being directly illuminated by the laser as opposed to being indirectly and inefficiently heated via a crucible.\cite{Smith:65, Hass:69}
The use of lasers also allows for small working distances, reducing the required source and chamber sizes when compared to other epitaxy techniques, and correspondingly, increasing material utilization.

Among the range of established thin film deposition methods,\cite{Capper2017} molecular beam epitaxy (MBE)\cite{MBE_Henini} may offer the advantage over PLD, e.g.\ for the epitaxial growth of complex oxides,\cite{Martin_OxideEpitaxy} in that it exploits adsorption-limited growth modes and the continuous variation of material composition by varying fluxes from individual sources.

\begin{figure}[!t]
	\centering
	\includegraphics[width = \linewidth]{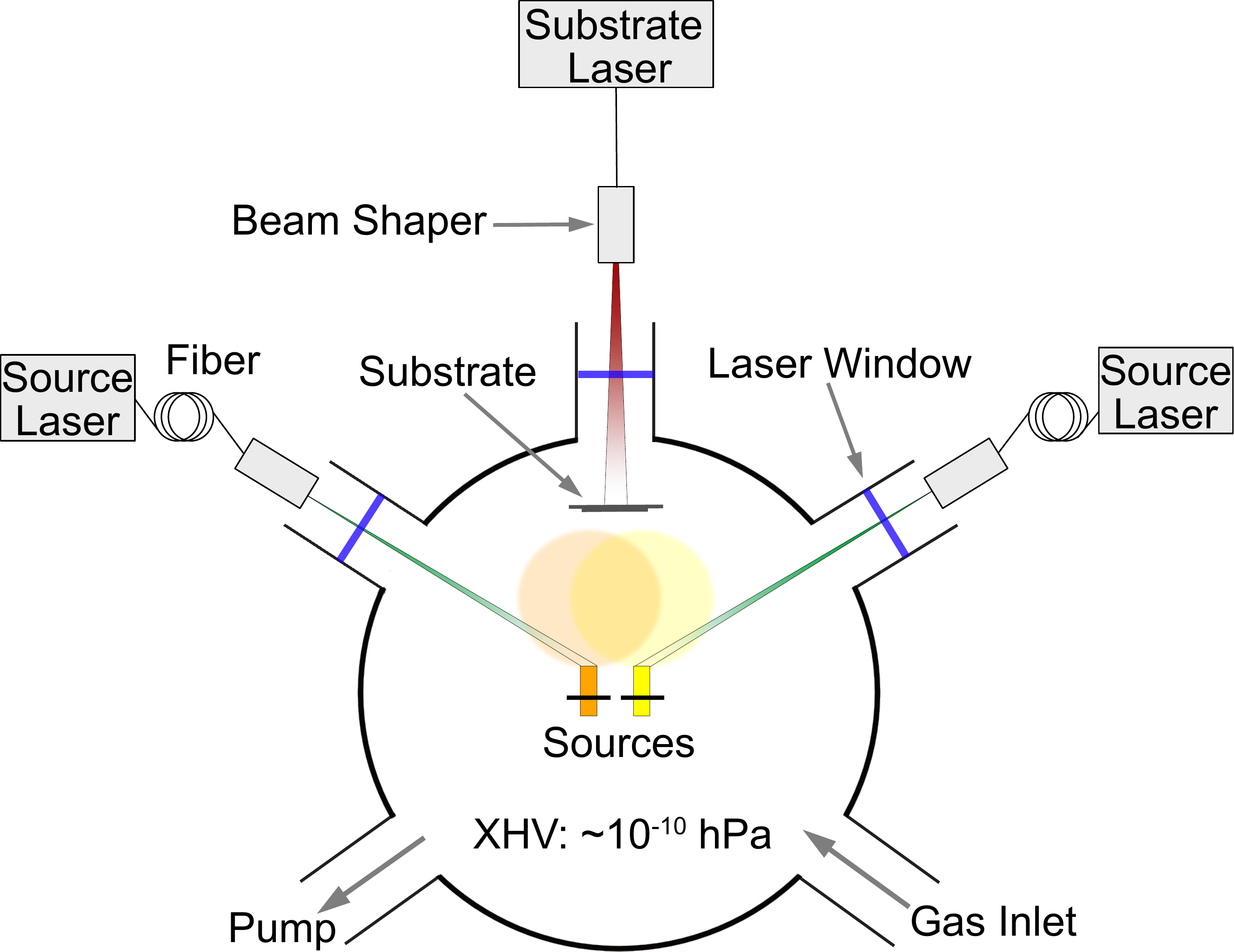}
	\caption{Diagram of the proposed TLE chamber for epitaxial growth. Without breaking the vacuum, both the substrate and the sources are exchangeable via the use of a load-lock. A gas inlet allows for a gaseous atmosphere to be present inside the chamber, which features an XHV background pressure. The pale colored regions show the approximate flux distributions from the sources.}
	\label{tlefull}
\end{figure}

Recently, high-power lasers have become sufficiently affordable to be practical for the generation of vapors from evaporating individual elemental sources. The laser heating of such elemental sources forms the basis for Thermal Laser Evaporation (TLE).\cite{braun}
TLE aims to use laser beams in order to evaporate a wide range of possible materials in the form of compact sources.
A co-evaporation from such sources would enable the epitaxial growth of compounds as films.
Substrate heating may be achieved by a heating laser\cite{subprep}, as shown in Fig.~\ref{tlefull}.
By varying the output power of each evaporation laser, a correspondingly variable composition may be achieved.
Investigating various extreme cases, we propose that successful evaporation and deposition can be achieved via TLE for all solid, non-radioactive elements.
Most of these experiments were performed using the same laser, with the same settings and parameters, varying only the output power.
 In addition, we demonstrate how free-standing sources can be utilized for nearly all of these elements and why the use of these sources is beneficial for epitaxial growth. 

\section*{Experimental Apparatus}

\begin{figure}
	\centering
	\includegraphics[width = \linewidth]{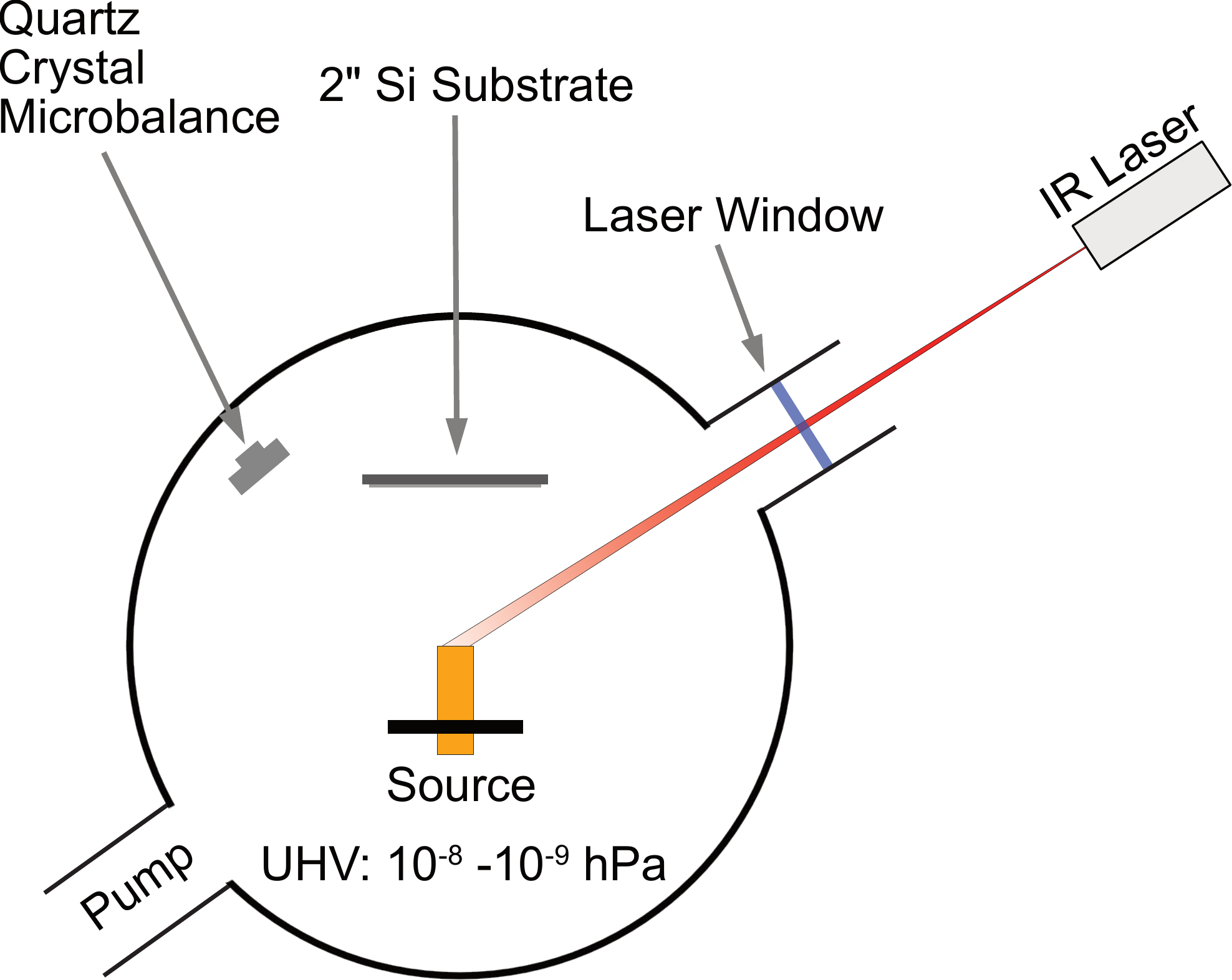}
	\caption{Diagram of the test TLE chamber. This test chamber operates without substrate heating.}
	\label{fig:my_label}
\end{figure}

The experiments outlined within this paper have been conducted using the simplified TLE chamber shown in Fig. \ref{fig:my_label}. This chamber had no substrate heater and only one source laser.
The chamber itself was not cooled and was pumped down to an ultra high vacuum (UHV) with internal pressures between 10$^{-8}$ and 10$^{-9}$\,hPa.
The cylindrical sources were free-standing with their long axes vertically oriented.
These sources rested on three Ta supports with a W-Re (type C) thermocouple measuring the temperature at the bottom of the source.
The use of this thermocouple provided a large measurement range with an upper limit at 2320\,$^\circ$C.
At the same time, it served as a safety control for the temperature of the bottom of the free-standing sources, in order to avoid complete melting of the source.
For elements that required a crucible, 12\,mm diameter \ce{Al_2O_3} or Ta crucibles were used, depending on the properties of the element in question.
The position of the source with respect to the laser beam could be translated in all three spatial dimensions via a manipulator on which the source was positioned.

For the dimensions of the elemental sources, we limited ourselves to cylinders with a maximum diameter of 12.7\,mm since materials with these dimensions are readily available from many suppliers.
The source heights along their axes were 8\,mm throughout unless explicitly stated.
Deposition rates greater than 0.1\,\AA/s were deemed to be great enough to confirm successful deposition of the corresponding element, based on typical growth rates that are suitably large for many film growth applications.\cite{mbegrowthrates}

For most of the elements tested, the source was irradiated by a $\lambda = 1030\,$nm fiber-coupled disk laser with a peak power of 2\,kW which was incident at 45$^\circ$ to its surface.\cite{Trumpf}
For seven of these elements (Bi, Ce, Sc, Cr, Ca, Yb and Mn), we switched to a $\lambda = 1070\,$nm fiber laser\cite{Trumpf} with a peak power of 500\,W, as it turned out that the required powers (see also Fig.~\ref{growthfit}) were consistently below 500\,W.
In all of these experiments, the distance between the source and the focal point of the laser was 62\,mm; this produced a laser spot size of approximately 1\,mm$^2$ with an approximately Gaussian beam profile.
The laser optics possessed a focal length of 500\,mm, which remained fixed for all the experiments.
Elemental films were deposited on a horizontal 2" Si substrate, centered 60\,mm vertically above the source.
Si substrates were used due to their low cost and ease of acquisition.
A quartz crystal microbalance (QCM) was applied to monitor the deposition from the source on the substrate.
The average growth rate was calculated after each ten-minute deposition test.
The growth rate was then checked by cleaving the Si wafers along their diameter and measuring the thickness of the grown films via scanning electron microscopy (SEM).
This comparison allowed for the calculation of the tooling factor to calibrate the deposition rate.
We estimate the uncertainty in the recorded growth rate to be less than the marker size within the relevant figures of this paper. 
The source was monitored during deposition with several CCD cameras. 

In the experiment, we controlled the heating in two ways: firstly, most of the experiments performed used power as an independent variable, wherein the laser power was kept constant during each deposition test and the source temperature was monitored via the thermocouple.
For these experiments, the average source temperature was measured over the time interval after the initial approach where the thermocouple temperature was relatively stable.
Secondly, the temperature of the bottom of the source could also be PID-controlled.
This control used the thermocouple temperature as input and the laser power as output which in turn ensured that the temperature of the bottom of the source remained at a predefined, fixed value for each deposition test.   

In order to minimize the effect of surface roughness and contaminants on the required power to achieve a specific growth rate, an initial deposition was performed with each source.
This melted or sublimated possibly existing small structures on the surface, yielding a steady-state surface morphology.
In the case of metals, the formation of a smooth surface could be easily observed by the corresponding increase in reflectivity. 

For all of the elements tested, no alteration to the laser parameters was required for either the disc laser or the fiber laser.
Only the output laser power was changed.
\begin{figure}
	\centering
	\includegraphics[width = \linewidth]{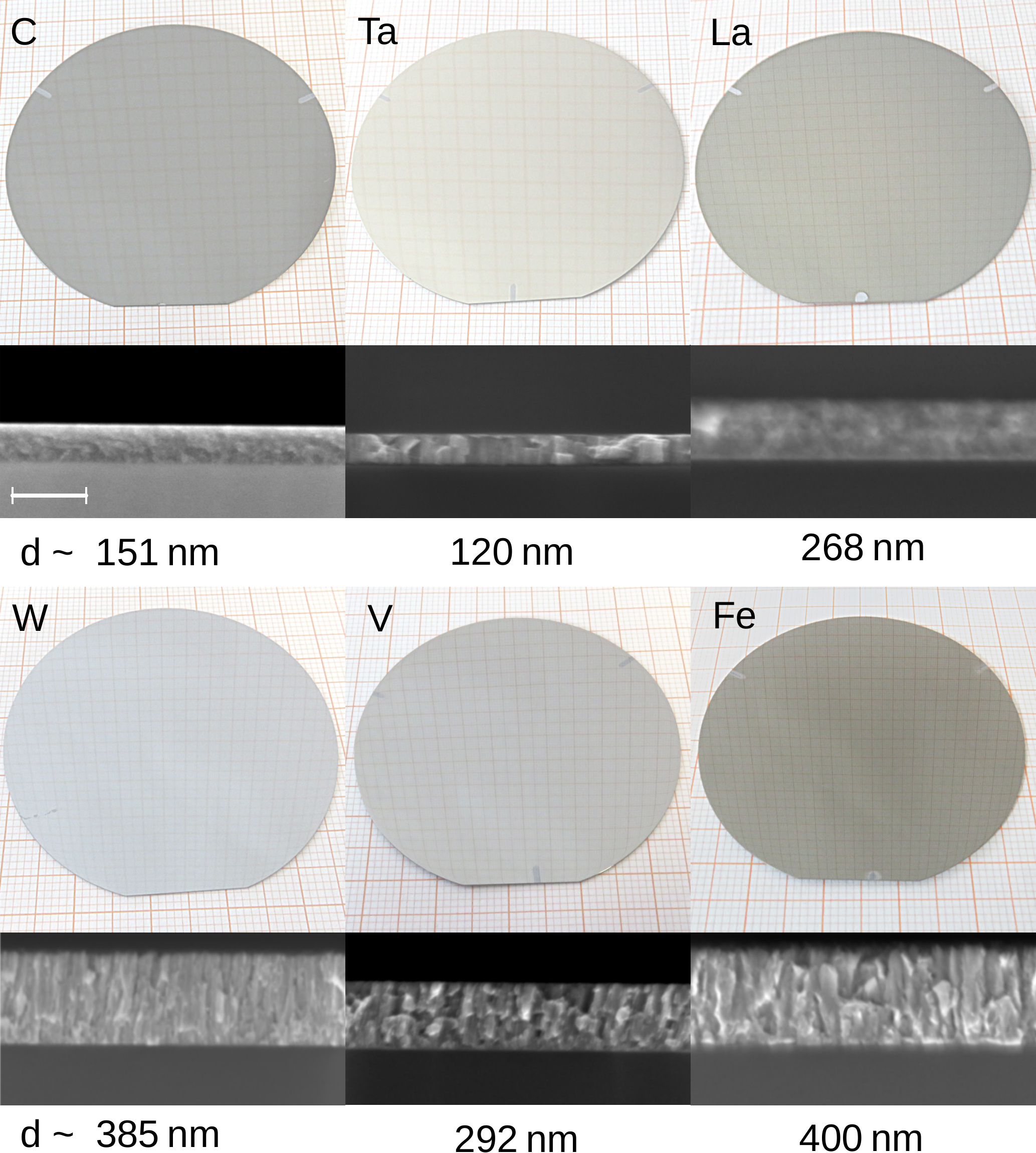}
	\caption{Photographs of Si substrates coated with various elements.
		The corresponding scanning electron microscopy images are shown below each wafer, along with the average thickness of the deposited films.
		The grid patterns appearing on the wafer surfaces result from optical reflections.
		The scale shown represents 300\,nm.}
	\label{wafer}
\end{figure}

\section*{Results}
Thin films with thicknesses greater than 100\,Å were grown with growth rates exceeding 0.1\,Å/s for all of the elements tested.
Elements from different parts of the periodic table could be easily deposited in a straightforward manner as thin films, with particular success for the metals (see also Fig. \ref{wafer}).
The films were dense, showing variations in thickness along the substrate between 5 and 15\,$\%$ for most of the elements tested, resulting in highly reflective films.
The results demonstrate the successful deposition of W, C and Ta using thermal evaporation induced via a CW laser.
Like other refractory metals, these elements have previously required e-beam evaporation to thermally deposit them due to their low vapor pressures and very high evaporation temperatures. \cite{Carbon,tungstenebeam,tantalumebeam}

\begin{figure}
	\centering
	\includegraphics[width = \linewidth]{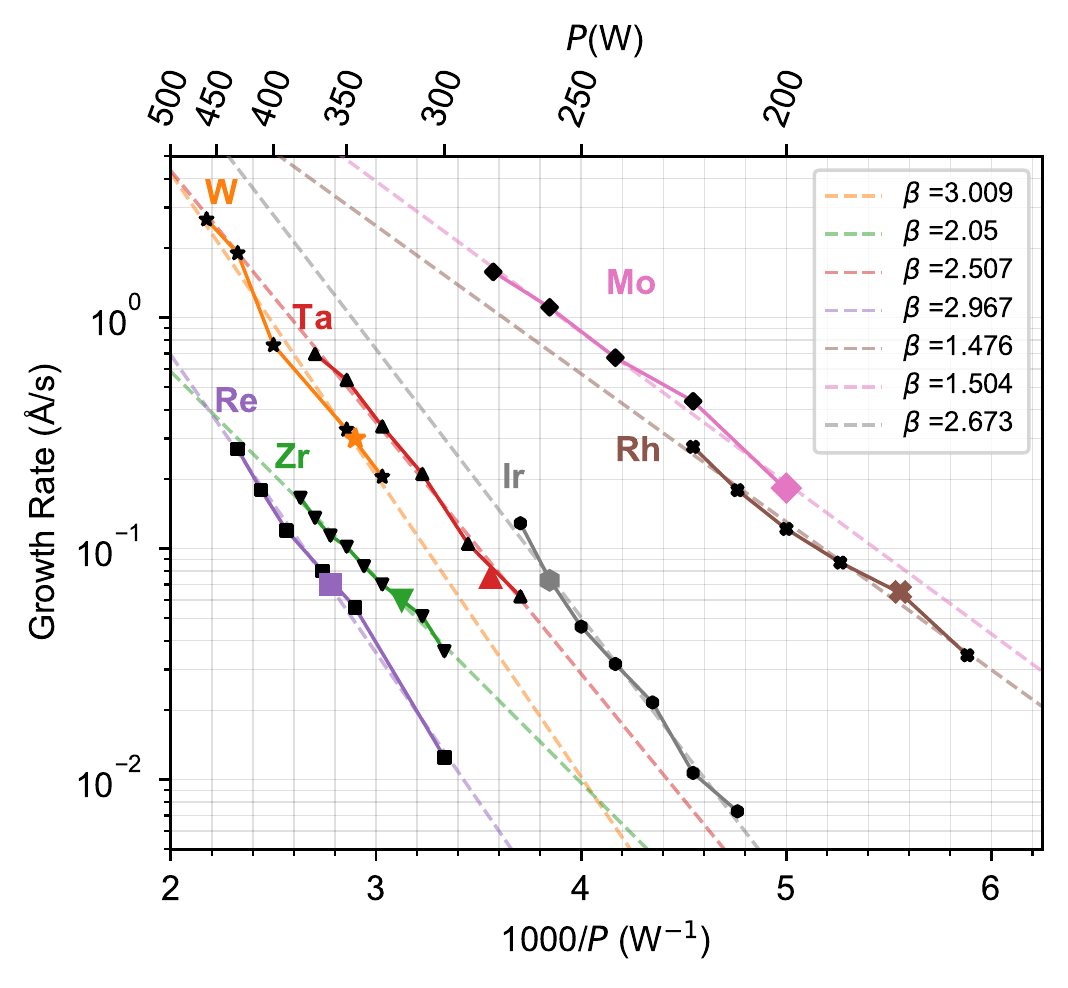}
	\caption{Plot of growth rate as a function of inverse incident power for a range of elements, accompanied by their associated fits to an Arrhenius-type dependence.
		The values of the proportionality constant, $\beta$ (see Eqn.\ \ref{equation1}) are indicated for each fit.
		For each element, the observed melting point is represented by a colored symbol.
		Connecting lines act as a guide to the eye. }
	\label{growthfit}
\end{figure}

Different laser power values were required for different elemental sources, as shown by Fig. \ref{growthfit}.
This figure shows the relation between growth rate and output laser power for various elements, accompanied by their associated fits to an Arrhenius-type dependence.
The observed melting points for each material are shown.
These served as a calibration point for the temperature of the source.
For most of the elements tested, deposition rates exceeding 1\,Å/s were possible with laser outputs less than 500\,W.
By optimizing the source dimensions, it is likely that this can be achieved for all elements in the chosen chamber geometry.
Assuming a wall-plug efficiency of 35\%\cite{Trumpf}, a maximum total power requirement of 1.4\,kW is obtained for a growth rate of 1\,Å/s.
This power requirement is similar to the corresponding requirements of many sputtering or e-beam systems. 
\begin{figure}[t]
	\centering
	\includegraphics[width = 0.7\linewidth]{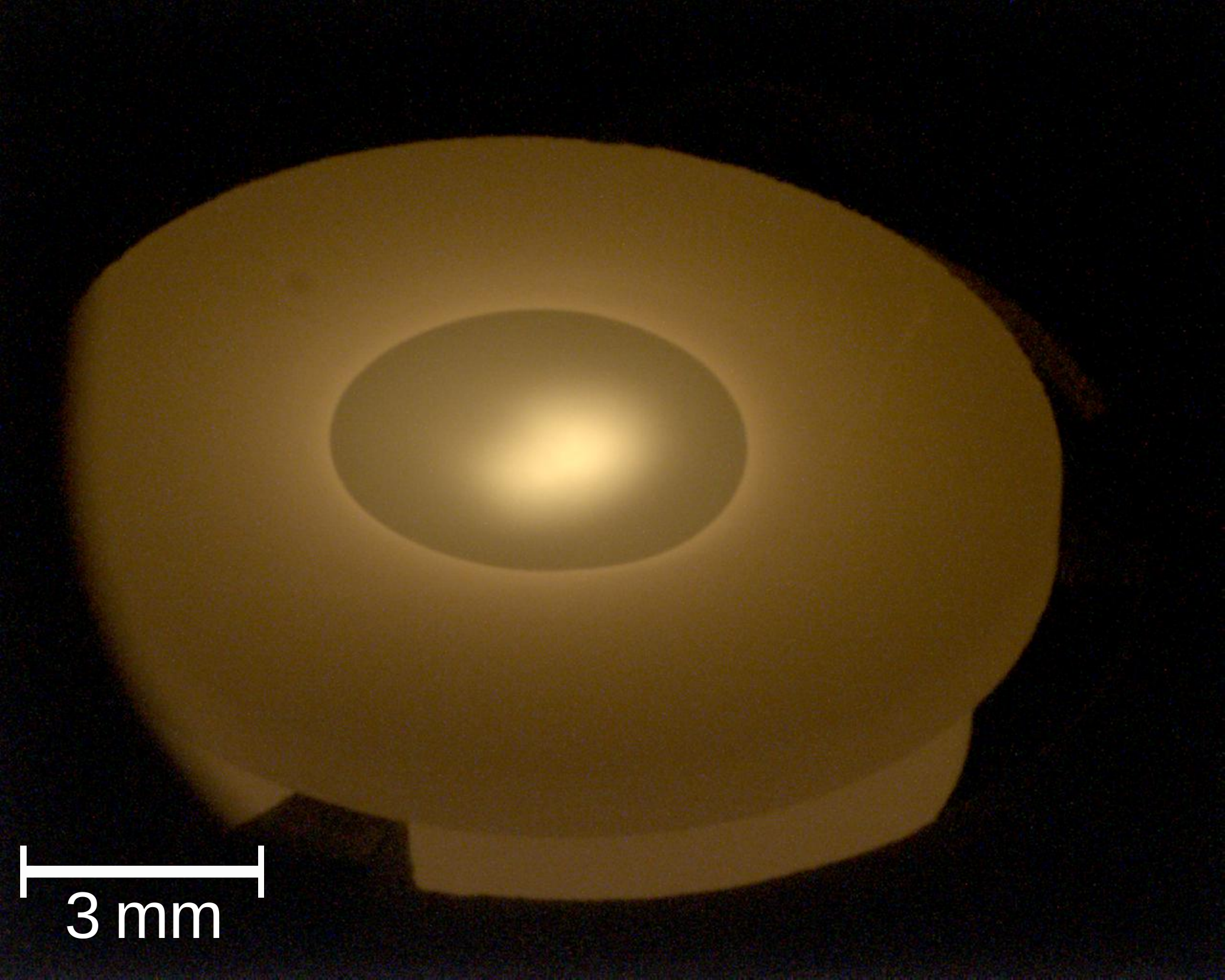}
	\caption{Photograph of a free-standing Si source (12.7\,mm in diameter and 4\,mm thick) during deposition.
		A liquid phase can clearly be seen on the surface of the source, indicating a surface temperature of at least 1687\,K.\cite{vapour}
		During this deposition test, the average deposition rate was 5.2\,Å/s, resulting from 400\,W supplied by the laser.}
	\label{source}
\end{figure}
\begin{figure}
	\centering
	\includegraphics[width = 0.7\linewidth]{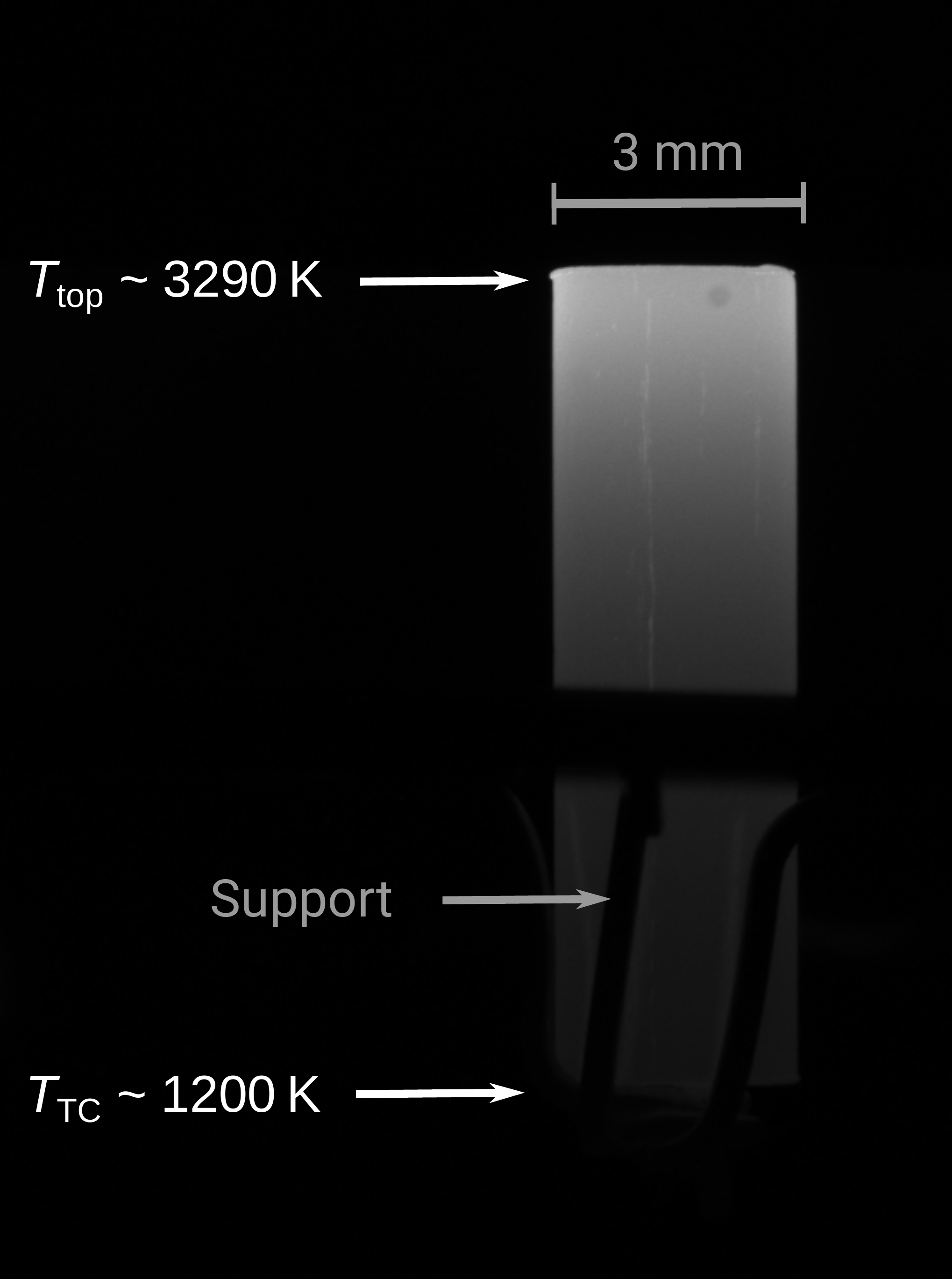}
	\caption{Photograph of the side of a free-standing Ta source (3\,mm in diameter and 10\,mm in height) during deposition.
		A clear intensity gradient can be seen along the (vertical) axis of the source, with the temperature difference between the top and the bottom of the source being $\sim$ 2000\,K as determined from the observation of liquid Ta on the top surface and a thermocouple measurement on the bottom surface.
		During this deposition test, the average deposition rate was 0.08\,Å/s, resulting from 280\,W supplied by the laser.}
	\label{Tasource}
\end{figure}

We were able to evaporate most of the elements from free-standing cylindrical sources, with localized laser heating of the source.
As a result, small regions of the source can be melted whilst the rest of the source remains solid, as shown in Fig. \ref{source}.
The source then acts as its own crucible, which offers major advantages.
The ductility of metals, combined with the absence of a mismatch in the thermal expansion between the 'crucible' and the 'contents' avoids the issue of crucible failure due to thermally induced strain.
The lack of a crucible also eliminates reactions between the source and the crucible, since the solid-liquid boundary remains inside the source.
This avoids contaminants generated from the crucible in the deposited film.

Owing to the strong absorption of light with a wavelength around 1000\,nm, the laser heating is not only highly localized via the small beam diameter, but also along the axis of the source.
The resulting absorption corresponds to a penetration depth on the order of 2\,nm for most metals with absorption coefficients; $\alpha \sim 10^{5}$\,cm$^{-1}$.
Due to the combination of heat loss via conduction and radiation, the high temperature region is found close to the irradiated surface of the source.
Figure \ref{Tasource} shows a side view of a Ta source during deposition, with a diameter of 3\,mm and a height of 10\,mm. 

In this case, a temperature difference of around 2000\,K was observed between the top surface and the thermocouple touching the bottom of the source.
Liquid Ta was present on the top surface, indicating the temperature exceeding 3290\,K\cite{vapour}, as denoted in Fig. \ref{Tasource}.
A temperature of approximately 1200\,K was measured by the thermocouple on the bottom of the source.
As a result of the localized heating from the source laser, the remainder of the source remained solid and the temperature of the bottom of the source did not exceed the range of operation for the thermocouple.
Depending on the thermal conductivity of the thermocouple contact to the source, inaccurate thermocouple temperature readings are expected at high temperatures.
Given that vapor pressure is highly dependent on temperature and proportional to the growth rate \cite{braun}, localized heating of the source also ensured that evaporation from the laser spot dominated the overall growth rate.
\begin{figure}
	\centering
	\includegraphics[width = 0.7\linewidth]{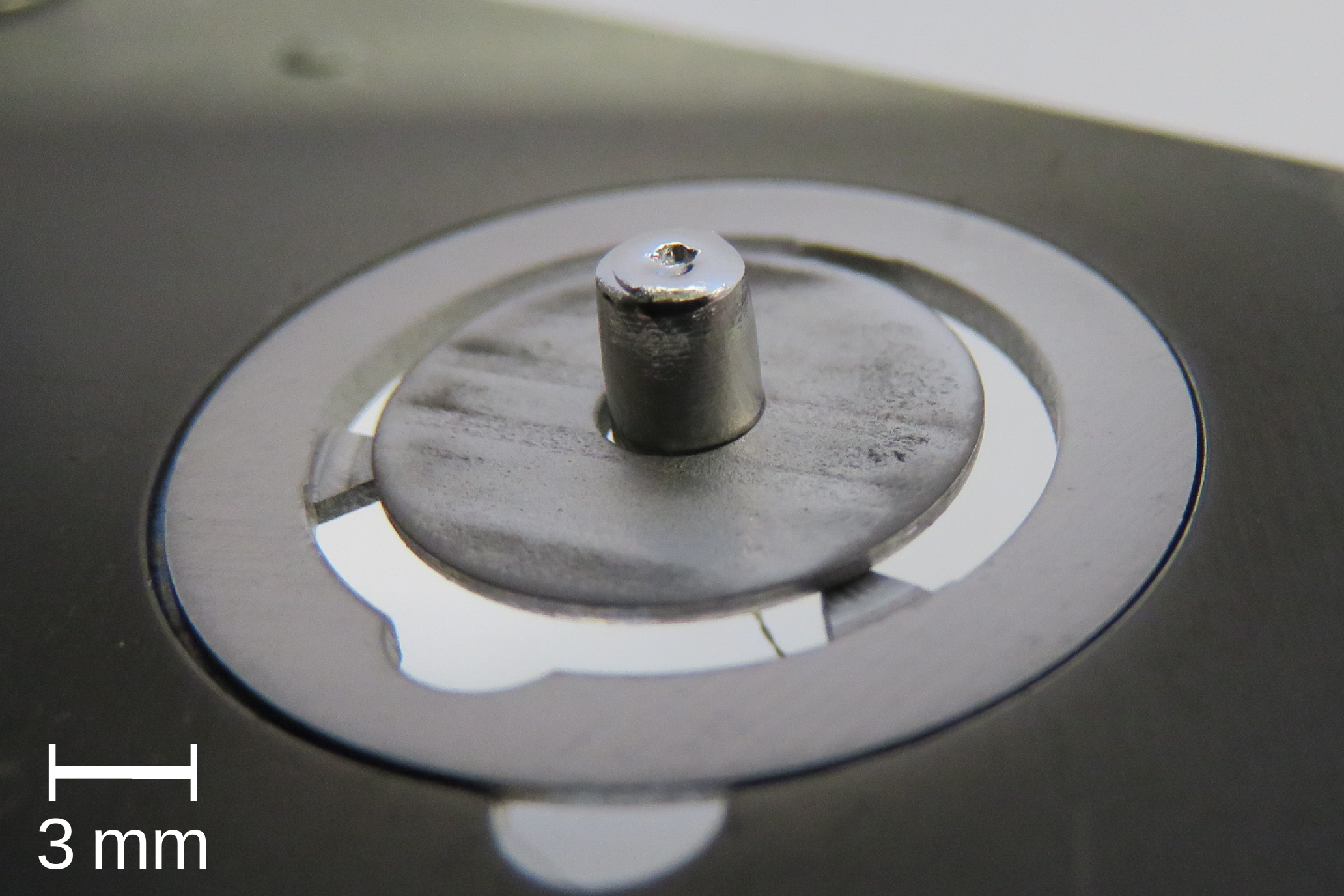}
	\caption{Photograph of a Ti source after approximately 300\,nm of Ti was deposited on the substrate. The surface has melted, generating a dip at the location of the laser spot.
		The structural integrity of the source has remained intact.}
	\label{Tisource}
\end{figure}

Pulsing the laser, while still remaining in the thermal evaporation regime, may further increase thermal gradients, allowing further concentration of the laser power to the evaporation region.\cite{prokhorov1990laser}
The experiments were nevertheless carried out under steady-state conditions, because the thermal transients at the onset and end of the pulses affect the reproducibility.
Further jitter in the pulse duration adversely affects the stability, adding additional noise to the steady-state (in)stability of the laser.

Even at high growth rates, the structural integrity of the free-standing source may be maintained.
For the case of a Ti source, 3\,mm in diameter (see  Fig. \ref{Tisource}), we observed that during deposition, the melt-pool was flat and could be increased in size until it was near the rim of the source, the remnant of which can still be seen in Fig. \ref{Tisource}.
However, upon cooling, a dip formed at the location of the laser spot, which we attribute to the contraction of Ti due to rapid cooling at the position of the laser spot, whilst solidifying inwards from the rim of the melt pool.
By reheating and melting the source, the flat surface was recovered.

Some elements, such as Ge and Pr, required a crucible in order to achieve growth rates greater than 0.1\,Å/s.
Whilst (in principle), a free standing source could be used for these elements too, our deliberate size restriction to below 12.7 x 8\,mm for the current experiments, did not allow a large enough source  for free-standing evaporation of these elements.

\section*{Discussion}
It is well-established within the field of film growth that vapor pressure, $p$, follows the Clausius-Clapeyron equation:
\begin{equation}
	p \sim e^{-\frac{\Delta H}{RT}},
\end{equation}
where $\Delta H$ is the enthalpy change between two desired phases and $R$ is the gas constant.
When this equation is plotted on an Arrhenius plot as a function of inverse temperature, a linear fit is obtained.
The growth rate (or evaporation rate) of an element at a specific temperature is proportional to the vapor pressure.\cite{Langmuir}
Since the vapor pressure is strongly dependent on temperature, the growth rate should be primarily dominated by evaporation from the laser spot.
However, the current experimental setup makes measuring temperature of the source at the laser spot at any temperature other than the melting point extremely difficult.
Therefore these experiments measured the growth rate as a function of the output laser power; a variable which is comparatively easy to measure.
Surprisingly, if one plots this growth rate data on an Arrhenius plot as a function of inverse laser power, as shown in Fig. \ref{growthfit}, the experimental data follows a straight line to a remarkable degree.

This result suggests that the growth rate as a function of inverse power follows an equation of the form:
\begin{equation}
\label{equation1}
 R_G \sim e^{-\frac{\beta}{P}},
\end{equation}
where $R_G$ denotes the growth rate, $P$ is the output laser power and $\beta$ the proportionality constant.
Given the prior relation between growth rate, $R_G$, and vapor pressure, $p$, an equivalence can be drawn between Eqn.\ \ref{equation1} and the Clausius-Clapeyron equation.
This leads to the interesting deduction that the relation between incident laser power and peak source temperature is approximately linear.
The proportionality constant from the $\beta$ values in Fig. \ref{growthfit} can be determined to:
\begin{equation}
\beta \sim \frac{\Delta H}{R a},
\end{equation}
where $a$ is the constant of proportionality between peak source temperature and output laser power.
We can use the Arrhenius fits obtained in Fig. \ref{growthfit} to obtain values of $a$ for each source material and geometry.
These values, along with their associated source diameters, are shown in Table I.

\begin{table}[h!]
	\begin{center}
	\begin{tabular}{| c| P{1.75cm}| c |}
		\hline
		Element & Source Diameter (mm) & $a$ (W/K) \\ 
		\hline
		W & 2 & 0.0127  \\ 
		\hline
		Zr & 12.7 & 0.005 \\
		\hline
		Ta & 3 & 0.0109  \\
		\hline
		Re & 3 & 0.012 \\
		\hline
		Rh & 5 & 0.004  \\ 
		\hline
		Mo & 2 & 0.007  \\ 
		\hline
		Ir & 3 & 0.008  \\
		\hline
	\end{tabular}
	\end{center}
\caption{Calculated $a$ values from the Arrhenius fits in Fig. \ref{growthfit}. The corresponding source diameters are also shown.}
\end{table}

An explicit model of this behaviour would have to account for phase changes, as these would affect the temperature distribution throughout the source resulting from convection inside the liquid phase, and the loss of energy due to vaporisation.

\begin{figure}
	\centering
	\includegraphics[width = \linewidth]{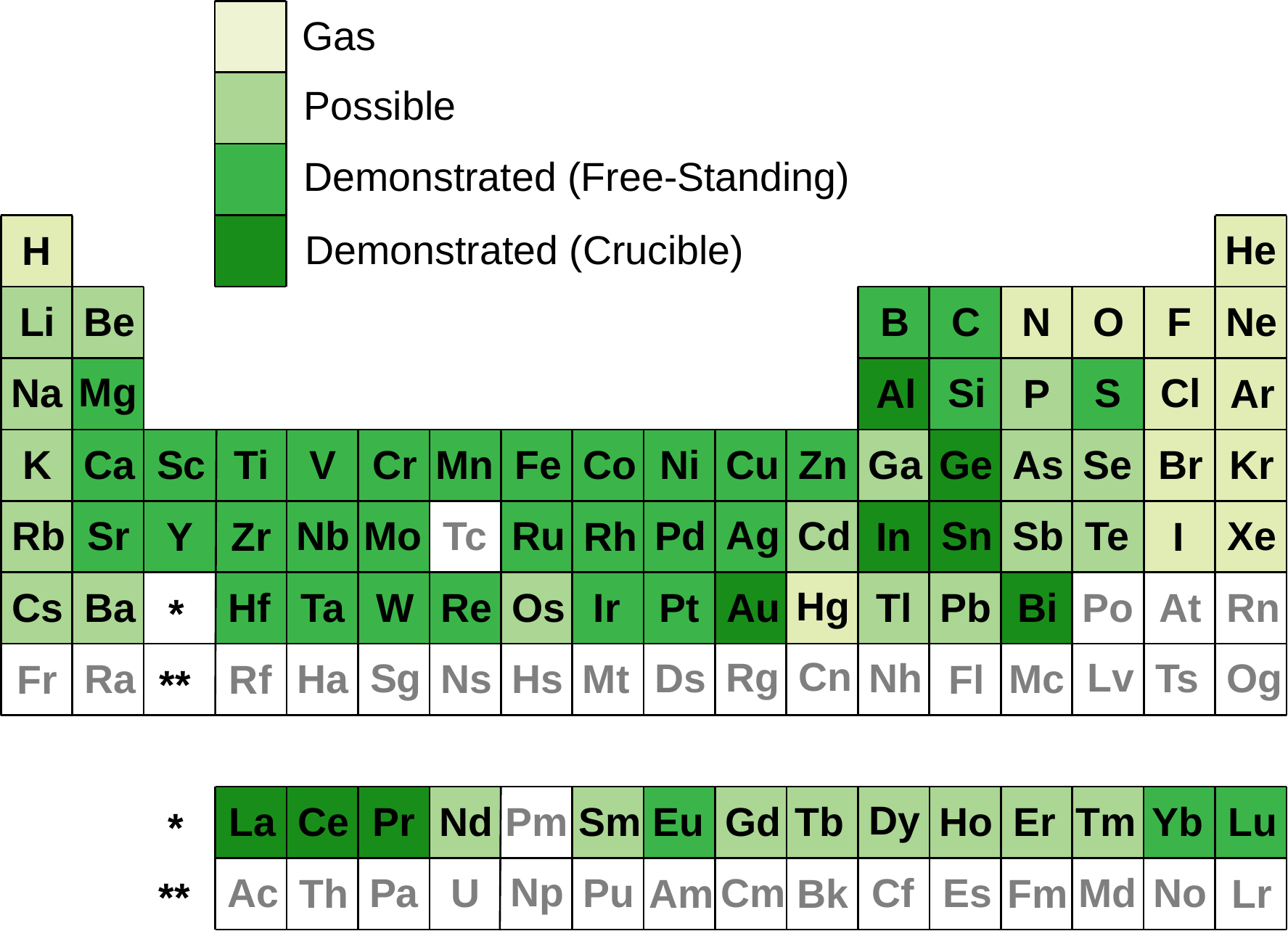}
	\caption{Periodic table showing in dark green the range of elements that in this work have been successfully deposited at substantial growth rates using free-standing sources and crucibles.
		Elements that have been grayed out are radioactive.}
	\label{periodic}
\end{figure}

The success we have achieved with our experimental results raises the question whether thermal laser evaporation can be applied to any solid, non-radioactive element in the periodic table.
To explore this question, we specifically investigated the growth of those elements with extreme growth parameters.
For example, being able to readily evaporate W, which has the lowest vapor pressure of any element, and a melting temperature of 3695\,K\cite{vapour} indicates that TLE is not limited by the evaporation temperature.
Similarly, successful evaporation of Cu, Ag and Au, all of which exhibit very high reflectivity values at 1030\,nm\cite{noble} indicates that reflectivity does not present a barrier to TLE.
High and low values of thermal conductivity also pose no issue, as we were able to successfully evaporate both Ag and S, respectively.
Therefore, we conclude that TLE can be applied to all solid, non-radioactive elements in the periodic table.
Our current experimental state is shown in Fig. \ref{periodic}, with the deposition of 43 elements being experimentally confirmed.

\section*{Conclusion}
We conclude that Thermal Laser Evaporation is a powerful technique to grow any solid, non-radioactive elements as thin films.
We have confirmed this experimentally for 43 elements by using the same laser beam geometry.
Due to the localized heating induced by the incident source laser, most of these elements could be deposited using free-standing sources.
Some of the investigated elements required a crucible, as determined by the chamber geometry.
Even for elements like W, C and Ta, deposition rates exceeding 1\,Å/s were achievable with laser powers less than 500\,W, with appropriately optimized source dimensions.

Whilst our focus here has been on film growth, one could easily foresee an application of this configuration in other areas of physics as well, such as the generation of atomic vapors for cold ion traps or for spectroscopic purposes.
\section*{Acknowledgments}
The authors thank Hans Boschker, Dong Yeong Kim and Sander Smink for many valuable discussions and support. In addition, the authors thank Fabian Felden, Ingo Hagel, Konrad Lazarus, Sabine Seiffert and Wolfgang Winter for technical assistance.


\begin{thebibliography}{21}
	
	\bibitem{Maiman_1960}
	T.~H. Maiman,
	"Stimulated {O}ptical {R}adiation in {R}uby,"
	\emph{Nature} \textbf{187}, 493--494, (1960).
	
	\bibitem{Smith:65}
	H.~M. Smith and A.~F. Turner,
	"Vacuum {D}eposited {T}hin {F}ilms using a {R}uby {L}aser,"
	\emph{Appl. Opt.} \textbf{4}, 147--148 (1965).

	\bibitem{Groh}
	G.~Groh,
	"Vacuum deposition of thin films by means of a \ce{CO_2} laser,"
	\emph{Journal of Applied Physics} \textbf{39}, 5804--5805 (1968).
	
	\bibitem{PLDWillmott}
	P.~R. Willmott and J.~R. Huber,
	"Pulsed laser vaporization and deposition,"
	\emph{Rev. Mod. Phys.} \textbf{72}, 315--328 (2000).
	
	\bibitem{Schwarz}
	H.~Schwarz and H.~A. Tourtellotte,
	"Vacuum {D}eposition by {H}igh-{E}nergy {L}aser with {E}mphasis on {B}arium {T}itanate {F}ilms,"
	\emph{Journal of Vacuum Science and Technology} \textbf{6}, 373--378 (1969).
	
	\bibitem{pld}
	D.~Dijkkamp, T.~Venkatesan, X.~D. Wu, S.~A. Shaheen, N.~Jisrawi, Y.~H. Min~Lee,
	W.~L. McLean, and M.~Croft,
	"Preparation of {Y}{B}a{C}u oxide superconductor thin films using pulsed laser evaporation from high {T}c bulk material,"
	\emph{Applied Physics Letters} \textbf{51}, 619--621 (1987).
	
	\bibitem{Hass:69}
	G.~Hass and J.~B. Ramsey,
	"Vacuum {D}eposition of {D}ielectric and {S}emiconductor {F}ilms by a \ce{CO_2} {L}aser,"
	\emph{Appl. Opt.} \textbf{8}, 1115--1118 (1969).
	
	\bibitem{Capper2017}
	P.~Capper, S.~Irvine, and T.~Joyce,
	\emph{Epitaxial Crystal Growth: Methods and Materials},
	Springer International Publishing, Cham, 2017,
	ISBN 978-3-319-48933-9.
	
	\bibitem{MBE_Henini}
	Mohamed Henini, editor,
	\emph{Molecular Beam Epitaxy},
	Elsevier, Oxford, 2013,
	ISBN 978-0-12-387839-7.
	
	\bibitem{Martin_OxideEpitaxy}
	L.~W. Martin, Y.~H. Chu, and R.~Ramesh,
	"Advances in the growth and characterization of magnetic, ferroelectric, and multiferroic oxide thin films,"
	\emph{Materials Science and Engineering R: Reports} \textbf{68}, 89--133 (2010).
	
	\bibitem{braun}
	W.~Braun and J.~Mannhart,
	"Film deposition by thermal laser evaporation,"
	\emph{AIP Advances} \textbf{9}, 085310 (2019).
	
	\bibitem{subprep}
	W.~Braun, M.~Jäger, G.~Laskin, P.~Ngabonziza, W.~Voesch, P.~Wittlich, and J.~Mannhart,
	"In situ thermal preparation of oxide surfaces,"
	\emph{APL Materials} \textbf{8}, 071112 (2020).
	
	\bibitem{mbegrowthrates}
	A.~Ritenour, A.~Khakifirooz, D.~A. Antoniadis, R.~Z. Lei, W.~Tsai, A.~Dimoulas, G.~Mavrou, and Y.~Panayiotatos,
	"Subnanometer-equivalent-oxide-thickness germanium p-metal-oxide-semiconductor field effect transistors fabricated using molecular-beam-deposited high-k/metal gate stack,"
	\emph{Applied Physics Letters} \textbf{88}, 132107 (2006).
	
	\bibitem{Trumpf}
	{Trumpf GmbH + Co. KG, Ditzingen, Germany}, Tru{D}isk 2000.
	\url{ https://www.trumpf.com/en_INT}.
	
	\bibitem{Carbon}
	G.C Fiaccabrino, X.-M Tang, N~Skinner, N.F {de Rooij}, and M~Koudelka-Hep,
	"Interdigitated microelectrode arrays based on sputtered carbon thin-films,"
	\emph{Sensors and Actuators B: Chemical} \textbf{35]}, 247 -- 254 (1996).
	Proceedings of the Sixth International Meeting on Chemical Sensors.
	
	\bibitem{tungstenebeam}
	J.~H. Souk, J.~F. O~Hanlon, and J.~Angillelo,
	"Characterization of electron beam deposited tungsten films on sapphire and silicon,"
	\emph{Journal of Vacuum Science \& Technology A} \textbf{3}, 2289--2292 (1985).
	
	\bibitem{tantalumebeam}
	M.~Cevro,
	"Ion-beam sputtering of \ce{(Ta_2O_5)_x}-\ce{(SiO_2)_{1−x}} composite thin films,"
	\emph{Thin Solid Films} \textbf{258}, 91 -- 103 (1995).
	
	\bibitem{vapour}
	{TU} {V}ienna {V}apor {P}ressure {C}alculator,
	\url{https://www.iap.tuwien.ac.at/www/surface/vapor_pressure}
	(accessed: 2020-07-09).
	
	\bibitem{prokhorov1990laser}
	A.M. Prokhorov, V.I. Konov, I.~Ursu, and N.~Mihailescu,
	\emph{Laser {H}eating of {M}etals},
	Series in Optics and Optoelectronics. Taylor \& Francis, 1990,
	ISBN 9780750300407.
	
	\bibitem{Langmuir}
	I.~Langmuir,
	"The vapor pressure of metallic tungsten,"
	\emph{Phys. Rev.} \textbf{2}, 329--342 (1913).
	
	\bibitem{noble}
	P.~B. Johnson and R.~W. Christy,
	"Optical {C}onstants of the {N}oble {M}etals,"
	\emph{Phys. Rev. B} \textbf{6}, 4370--4379 (1972).
	
\end{thebibliography}
\end{document}